\documentstyle[aps,twocolumn,epsf,epsfig]{revtex}
\begin{document}
\draft
\title{Number--phase uncertainty relations: 
verification by homodyning}
\author{T. Opatrn\'{y}$^{1,2}$,
M. Dakna$^{1}$,    
and D.--G. Welsch$^{1}$}
\address{$^{1}$ Friedrich-Schiller-Universit\"at  Jena,\
Theoretisch-Physikalisches Institut\\
Max-Wien-Platz 1, 07743 Jena, Germany \\
$^{2}$ Department of Theoretical Physics, Palack\'{y} University \\
Svobody 26, 77146 Olomouc, Czech Republic}
\date{April 18, 1997}
\maketitle
\begin{abstract}
It is shown that fundamental uncertainty relations between photon number
and canonical phase of a single-mode optical field can be verified by
means of balanced homodyne measurement. All the relevant quantities can be
sampled directly from the measured phase-dependent quadrature distribution.
\end{abstract}



Although the problem of number--phase uncertainty has widely
been studied, there has been no direct experimental verification
of fundamental uncertainty relations (URs). What can be the best way of
doing that? A powerful and perhaps ultimate method for measuring
the quantum statistics of traveling optical fields has been balanced
homodyne detection. The quantity that is directly measured is
the probability distribution $p(x,\vartheta)$ of the phase-dependent 
quadrature $\hat{x}(\vartheta)$ $\!=$ $\!2^{-1/2}(\hat{a} e^{-i\vartheta}$ 
$\!+$ $\!\hat{a}^{\dag} e^{i \vartheta})$, where $\hat{a}$ ($\hat{a}^{\dag}$) 
is the bosonic annihilation (creation) operator of the (single-mode)
signal field and $\vartheta$ corresponds to the
local-oscillator phase. It has been shown that $p(x,\vartheta)$ for all 
phases $\vartheta$ in a $\pi$ interval contains all knowable information 
about the quantum state of the signal field and can be used to reconstruct 
the Wigner function applying inverse Radon transformation techniques 
\cite{VogelRisken}.

Since the Wigner function is a full description of the quantum state,
it can  be used to calculate other important features
of the field, such as the photon-number and phase statistics 
and their associated URs\cite{Smithey}. The inverse Radon transform
requires a three-fold integration of the measured data, and
the calculation of the density matrix in the Fock basis can then be
accomplished with two integrals. One summation eventually yields the
photon-number moments, and one sum and one integral must be
performed to obtain the (canonical) phase moments. Hence,
six- and seven-fold transformations of the recorded data are
required for UR verification at least.
Of course, large amount of data manipulation accumulates various errors 
and the physical nature of the uncertainties
becomes less transparent.

Recent progress has offered new possibilities of determining the 
photon-number statistics in a more direct way avoiding the detour
via the Wigner function. It has been shown that both the density-matrix
elements $\varrho_{nn'}$ in the Fock basis \cite{Ariano}
and the moments and correlations $\langle \hat{a}^{\dagger k}
\hat{a}^{k'}\rangle$ \cite{Richter2} can directly be sampled
from the recorded data according to a two-fold integral transform
\begin{eqnarray}
\label{n1}
{\cal A} = \int_{2\pi} d\vartheta \int_{-\infty}^{\infty} dx \,
K_{\cal A}(x,\vartheta) p(x,\vartheta) .
\end{eqnarray}
Here, ${\cal A}$ is the quantity that is to be determined
and $ K_{\cal A}(x,\vartheta)$ is the corresponding
integral kernel (sampling function). In particular, ${\cal A}$ can
be identified with $p_{n}$ $\!=$ $\!\varrho_{nn}$ or $\langle\hat{n}^{k}
\rangle$ $\!=$ $\langle(\hat{a}^{\dagger}\hat{a})^{k}\rangle$ in order
to determine the photon-number distribution or the photon-number moments.

In contrast to the photon number, the phase has remained a troublesome 
variable. Recently it has been suggested to measure the canonical
phase distribution asymptotically by replacing the local-oscillator
with a reference mode prepared in so-called reciprocal binomial states --
a method that is state dependent and hardly realizable at present
\cite{Barnett}. It has also
been suggested to use balanced homodyning for asymptotically sampling  
phase distributions \cite{Moh}. Again, the method is state
dependent and the exact phase distributions can be obtained
only in the sense of a limiting process. Moreover, with increasing 
(photon-number) excitation of the state under consideration an increasing 
number of highly oscillating functions must be summed up to obtain the 
corresponding integral kernels and high-accuracy sampling of the phase 
distributions from the recorded data becomes very difficult. 
The problem is similar to that of a calculation of phase quantities
from the sampled density-matrix elements in the Fock basis. 
With increasing excitation of the state a larger number of 
density-matrix elements must be sampled, so that the accumulated error 
may eventually dominate the result. 

In what follows we show that fundamental number--phase URs that are
based on exponential phase measures can be verified directly from the 
homodyne data according to Eq.~(\ref{n1}), without making a detour via 
other quantities and without making any assumptions and approximations 
with regard to the state. Let us consider a phase distribution
$p(\varphi)$ and its exponential phase moments defined by 
\begin{equation}
\Psi_{k} = \int_{2\pi}d\varphi \, e^{ik\varphi} p(\varphi) .
\label{00}
\end{equation}
In particular for the canonical phase the relation 
$\Psi_{k}$ $\!=$ $\langle\hat{E}^{k}\rangle$ is valid, where
the exponential phase operator $\hat E$ is given by 
$\hat E$ $\!=$ $\!(\hat n +1)^{-1/2}\hat a$ (for the canonical phase
and the exponential phase operator, see, e.g., \cite{Car}). 
In order to define a mean phase $\bar{\varphi}$ that is independent
of the chosen phase window, the first-order exponential phase moment
has been introduced in the definition, $\bar \varphi$
$\!=$ $\!{\rm arg}\Psi_{1}$ $\!=$ $\!{\rm arg}\langle \hat E \rangle$,
and it has been used to define phase uncertainty measures, such as
\begin{eqnarray}
 \label{n2}
 \Delta \varphi = \arccos |\Psi_{1} |
= \arccos |\langle \hat E \rangle| 
\end{eqnarray}
\cite{Tom1}, the Bandilla-Paul dispersion $\sigma^{2}_{\rm BP}$ 
\cite{Bandilla}, and the Holevo dispersion $\sigma^{2}_{\rm H}$ 
\cite{Holevo},
\begin{equation}
\sigma_{\rm BP} = \sin \Delta \varphi,
\quad 
\sigma_{\rm H} = \tan \Delta \varphi.
\label{01}
\end{equation} 
It can then be proven that the UR
\begin{eqnarray}
\label{u1}
\Delta n \tan \Delta \varphi \ge {\textstyle\frac{1}{2}}
\end{eqnarray}
is valid \cite{Tom1},
which is equivalent to the Holevo UR \cite{Holevo}
\begin{equation}
(\Delta n)^{2} \sigma^{2}_{\rm H} \ge {\textstyle\frac{1}{4}}
\label{02}
\end{equation}
[$(\Delta n)^{2}$ $\!=$ $\!\langle \hat n^{2} \rangle$ $\!-$
$\!\langle \hat n \rangle ^{2}$]. Although these URs are exact, they are, 
in a sense, weak. This means that they also allow for such values of 
$\Delta n$ and $\Delta \varphi$ for which no state exists (for more 
specific URs and the corresponding minimizing states, see 
\cite{BaPaRitze,Luks1}). Further, URs that are based on the 
Susskind--Glogower trigonometric operators
$\hat C$ $\!=$ $\!\frac{1}{2}(\hat E + \hat E^{\dag})$ and
$\hat S$ $\!=$ $\!\frac{1}{2i}(\hat E - \hat E^{\dag})$ 
have also been studied \cite{Car,Luks1,Kit},
\begin{equation}
 \label{n3}
 \Delta n \Delta C \ge {\textstyle\frac{1}{2}} 
 |\langle \hat S \rangle |, 
\quad
 \Delta n \Delta S \ge {\textstyle\frac{1}{2}} 
 |\langle \hat C \rangle |, \\
 \label{n3b}
\end{equation}
\begin{equation} 
 \Delta S \Delta C \ge {\textstyle\frac{1}{2}} \varrho_{00} 
\label{n3c} 
\end{equation} 
[$(\Delta C)^{2}$ $\!=$ $\!\langle \hat C^{2} \rangle$ $-$ $\langle \hat C
\rangle ^{2}$, and $(\Delta S)^{2}$ accordingly]. Note that the squares of
the trigonometric operators can be written as
\begin{eqnarray}
 \label{n4}
 \hat C^{2} = {\textstyle\frac{1}{2}} \hat 1 
 +  {\textstyle\frac{1}{4}} \left(
 \hat E^{2} + \hat E^{\dag 2} \right) 
 - {\textstyle\frac{1}{4}} |0\rangle \langle 0|, \\
 \hat S^{2} = {\textstyle\frac{1}{2}} \hat 1 
 - {\textstyle\frac{1}{4}} \left(
 \hat E^{2} + \hat E^{\dag 2} \right) 
 - {\textstyle\frac{1}{4}} |0\rangle \langle 0| .
\end{eqnarray}
   From the definitions of $\Delta \varphi$, $\Delta C$
and $\Delta S$ we see that for measuring them it is sufficient to measure
the exponential phase moments $\Psi_{1}$ $\!=$ $\langle \hat E \rangle$ 
and $\Psi_{2}$ $\!=$ $\langle \hat E^{2} \rangle$
and the density-matrix element $\varrho _{00}$ $\!=$ 
$\!\langle |0\rangle \langle 0 | \rangle$. The determination of
the photon-number uncertainty requires measurement of 
$\langle \hat n \rangle$ and $\langle \hat n^{2} \rangle$.

It is well known that $\varrho _{00}$, $\langle \hat n \rangle$, and 
$\langle \hat n^{2} \rangle$ can directly be sampled from
the data recorded in balanced homodyning applying Eq.~(\ref{n1}).
The kernel $K_{00}(x,\vartheta)$ for $\varrho _{00}$
can be taken from \cite{Ariano}, $K_{00}(x,\vartheta)$ $\!=$
$\!\pi^{-1}$ $\!\Phi (1,\frac{1}{2},-x^{2})$,
where $\Phi (a,b,z)$ is the confluent hypergeometric function.
The kernels $K_{n}(x,\vartheta)$ and $K_{n^{2}}(x,\vartheta)$
for $\langle \hat n \rangle$ and $\langle \hat n^{2} \rangle$, respectively,
can simply be obtained from the sampling formula for the normally
ordered moments and correlations of bosonic operators \cite{Richter2},
\begin{eqnarray}
\lefteqn{
\langle \hat{a}^{\dagger n}\hat{a}^m\rangle
}
\nonumber \\ && \hspace{2ex}
= \int_{2\pi}\! \! d \vartheta  \, e^{i(n-m)\vartheta}
\!\!\int_{-\infty}^{\infty}\! \! dx\, 
\frac{{\rm H}_{n+m}(x)}
{2\pi\sqrt{2^{n+m}} 
{n+m\choose m}} \, p(x,\vartheta),
\label{Ma1}
\end{eqnarray}
from which we find that  
\begin{eqnarray}
\label{nph1}
K_{n}(x,\vartheta) = {\textstyle \frac{1}{2\pi}} 
\left( x^{2} -{\textstyle \frac{1}{2}} \right) ,
\\
\label{nph2}
K_{n^{2}}(x,\vartheta) = {\textstyle \frac{1}{2\pi}} 
\left( {\textstyle \frac{2}{3}} x^{4}
- x^{2} \right) .
\end{eqnarray}
Note that for  large $|x|$ the leading terms in these expressions
determine the kernels for the energy moments in the classical limit.

Let us now turn to the problem of direct sampling of the
exponential phase moments of the canonical phase. Provided that the
corresponding integral kernels exist, their asymptotic behavior for
large $|x|$ can be obtained from considering the classical limit.
Since in classical physics the phase probability distribution
$p(\varphi)$ can be obtained from the phase-space probability distribution
$W(r,\varphi)$ according to $p(\varphi)$ $\!=$ $\!\int_{0}^{\infty} rdr$
$\!W(r,\varphi)$, the exponential phase moments $\Psi_{k}$ can be given by
\begin{eqnarray}
\label{cl1}
\Psi_{k} = \int_{2\pi} \int_{0}^{\infty}
\! rdr \, W(r,\varphi) e^{ik\varphi} .
\end{eqnarray}
Further, the quadrature probability $p(x,\vartheta)$ is given by the
Radon transform
\begin{eqnarray}
\label{cl2}
p(x,\vartheta) = \int_{2\pi}\! \! d\varphi \int_{0}^{\infty} \! \! r dr
\,W(r,\varphi)\,
\delta\! \left[ x\!-\!r \cos (\vartheta \! -\! \varphi) \right] .
\end{eqnarray}
Let us now assume that $\Psi_{k}$ can be related to $p(x,\vartheta)$
according to Eq.~(\ref{n1}) [${\cal A}$ $\!\to$ $\!\Psi_{k}$ and
$K_{\cal A}(x,\vartheta)$ $\!\to$ $\!K_{k}(x,\vartheta)$].
Substituting in this equation for $p(x,\vartheta)$ the result of
Eq.~(\ref{cl2}) and comparing with Eq.~(\ref{cl1}), we observe
that $K_{k}(x,\vartheta)$ can be written as
$K_{k}(x,\vartheta)$ $\!=$ $\!e^{ik\vartheta} K_{k}(x)$, where
$K_{k}(x)$ must satisfy the integral equation
\begin{eqnarray}
\label{cl3}
\int_{2\pi} d\varphi \, e^{ik\varphi} K_{k}(r \cos \varphi) = 1
\end{eqnarray}
for any $r$ $\!>$ $\!0$. From Eq.~(\ref{cl3}) we can see that
$K_{k}(x)$ is not uniquely defined. First, any function of parity
$(-1)^{k+1}$ can be added to $K_{k}(x)$ without changing the integral.
Second, any polynomial of a degree less than $k$ can also be added to
$K_{k}(x)$.
As can be verified, Eq.~(\ref{cl3}) is solved using the functions
\begin{eqnarray}
\label{cl4}
K_{2m\!+\!1}(x) = {\textstyle\frac{1}{4}}
(-1)^{m} (2m+1)\, {\rm sign} (x)
\end{eqnarray}
and
\begin{eqnarray}
\label{cl5}
K_{2m}(x) = \pi^{-1} (-1)^{m+1} m \log |x|
\end{eqnarray}
($m$ $\!=$ $\!0,1,2,\ldots$).
It is worth noting that since Eq.~(\ref{cl2}) is also valid when
$W(r,\varphi)$ is the quantum-mechanical Wigner function,
using in homodyne detection the kernels (\ref{cl4}) and (\ref{cl5})
over the whole $x$ axis would yield the exponential phase
moments of the phase quasi-probability distribution defined by
the radially integrated Wigner function.

With regard to the canonical phase, the kernels (\ref{cl4}) and (\ref{cl5})
are of course valid only for large $|x|$. To obtain them for arbitrary
$|x|$, we recall that in quantum physics $\Psi_{k}$ can be given by
\begin{equation}
\Psi_{k} = \langle\hat{E}^{k}\rangle =
\sum_{n=0}^{\infty} \varrho_{n+k\,n}
\label{03}
\end{equation}
in place of Eq.~(\ref{cl1}). Expressing $p(x,\vartheta)$ in terms of
the density-matrix elements $\varrho_{nn'}$ as
\begin{eqnarray}
\label{04}
p(x,\vartheta) = \sum_{n,n'\!=\!0}^{\infty} e^{i(n'-n)\vartheta}
\psi_{n}(x) \psi_{n'}(x) \varrho_{nn'}  ,
\end{eqnarray}
and assuming that Eq.~(\ref{n1}) applies to
$\Psi_{k}$, we again find that $K_{k}(x,\vartheta)$ $\!=$
$\!e^{ik\vartheta} K_{k}(x)$, but in place of Eq.~(\ref{cl3})
$K_{k}(x)$ must now satisfy the integral equation
\begin{eqnarray}
\label{qua1}
2\pi \int_{-\infty}^{\infty} \! dx \
K_{k}(x) \psi_{n}(x)  \psi_{n\!+\!k}(x) = 1
\end{eqnarray}
for every $n$ $\!=$ $\!0,1,2,\dots\,.$ Here,
$\psi_{n}(x)$ $\!=$ $\!(2^{n}n!\sqrt{\pi})^{-1/2}$
$\!{\rm exp}(-x^{2}/2)$ $\!{\rm H}_{n}(x)$ are the energy eigenfunctions
of a harmonic oscillator, with  H$_{n}(x)$ being the Hermite polynomials.
   From Eq.~(\ref{qua1}) and the properties of the Hermite polynomials
\cite{Prudnikov1} the same arbitrariness in the determination of
$K_{k}(x)$ as in the classical limit is found. To derive an
explicit expression, we use the expansion
\begin{eqnarray}
\lefteqn{
\hat E^{k} =\sum_{n=0}^\infty
:\frac{\hat{a}^\dagger\,^n\exp(-\hat{a}^\dagger\hat{a})\hat{a}^{n+k}}
{\sqrt{n!(n+k)!}}:
}
\nonumber \\ && \hspace{1ex}=\,
\sum_{n=0}^\infty\sum_{m=0}^\infty
\frac{1}{\sqrt{n!(n+k)!}}\frac{(-1)^m}{m!}
\hat{a}^{\dagger n+m}\hat{a}^{n+m+k}
\label{Ma3}
\end{eqnarray}
($:\ :$ introduces normal order)
and apply Eq.~(\ref{Ma1}) in order to represent
$\langle\hat{E}^{k}\rangle$ in the form of Eq.~(\ref{n1}).
Provided that all the correlations
$\langle \hat{a}^{\dagger n}\hat{a}^{n+k}\rangle$ exist, we derive
\begin{equation}
K_{k}(x) =  (2\pi)^{-1} \sum_{l=0}^{\infty}
C_l^{(k)} {\rm H}_{2l+k}(x),
\label{05}
\end{equation}
where
\begin{eqnarray}
C_l^{(k)}\!=\! \frac{(l+k)!}{2^{l+(k/2)}(2l\!+\!k)!}
\sum_{n=0}^l{\l\choose n}\! \frac{(-1)^{l\!-\!n}}
{\sqrt{(n\!+\!1)\dots (n\!+\!k)}}  \, .
\label{Ma5}
\end{eqnarray}
In particular, the kernels for sampling the first two moments can be
rewritten  as, on using standard relations \cite{Prudnikov1},
\begin{eqnarray}
\label{ker1}
K_{1}(x) = \pi ^{-3/2} 
x \int_{0}^{\infty} \!\! \frac{dt}{\sqrt{t}\,{\rm cosh}^{2}t}
\,\Phi\!\left( 2, {\textstyle\frac{3}{2}}, -x^2 {\rm tanh}\,t \right)
\end{eqnarray}
(also see \cite{DAriano}), and (apart from an irrelevant constant)
\begin{eqnarray}
\label{ker2}
K_{2}(x) = \frac{1}{2\pi}
\int_{0}^{\infty} \! dt \,
\frac{-{\rm I}_{0}(t)}{{\rm cosh}^{2} t \, {\rm sinh}\, t}
\,\Phi\!\left( 2, {\textstyle\frac{1}{2}}, -x^{2} {\rm tanh} \,t \right)
\end{eqnarray}
(I$_{0}(t)$ is the modified Bessel function).
Finally, it can be proven that the kernels exist and
satisfy the condition (\ref{qua1}), i.e., we have found
solutions even when the assumption of finite moments fails and
Eq.~(\ref{Ma1}) cannot be used.

\begin{figure}
\noindent
\vspace*{-2.cm}
\hspace*{0.5cm}
\begin{minipage}[b]{1.1\linewidth}
\unitlength1.2cm
\begin{picture}(4,7)
\epsfysize=6cm
\epsfbox{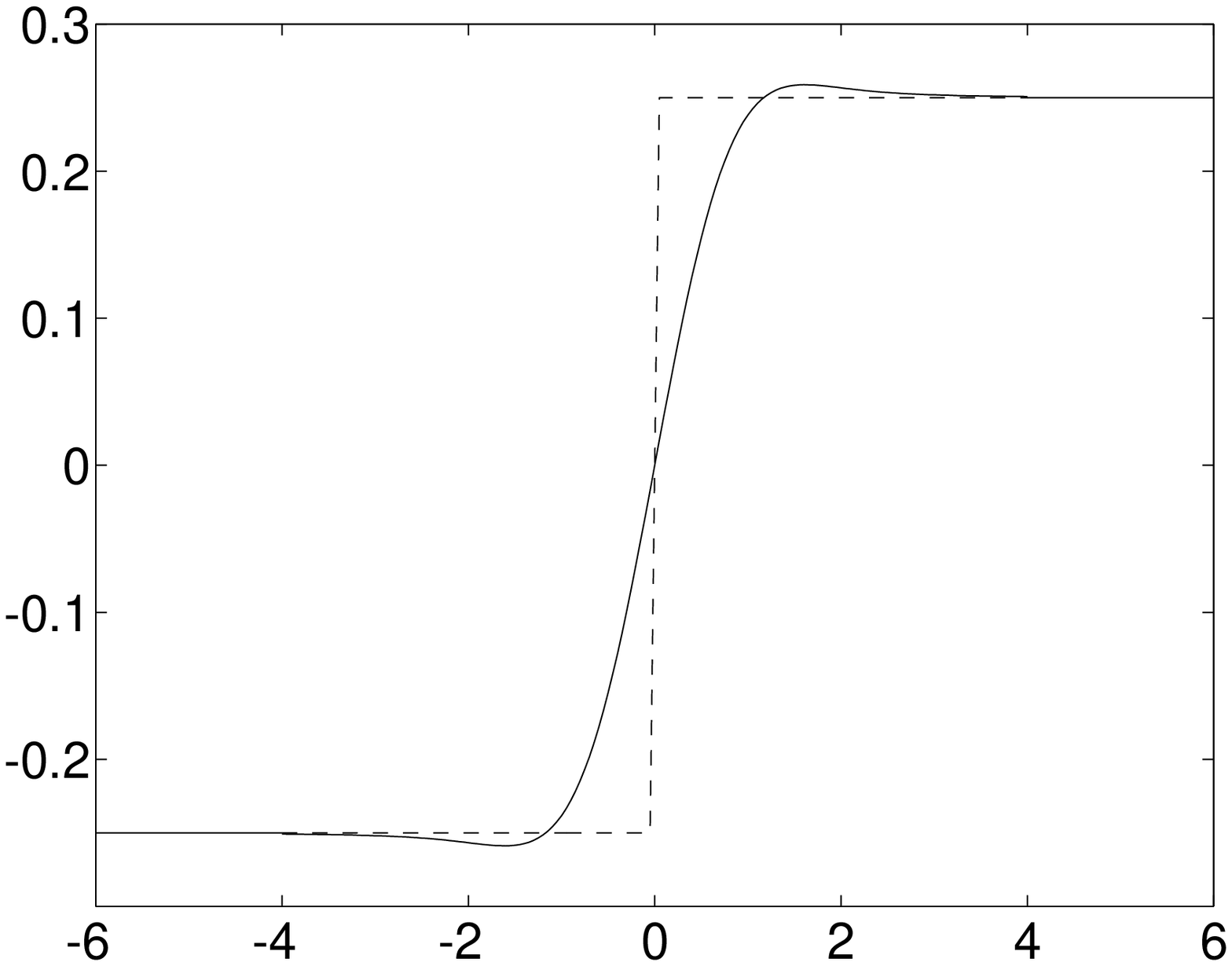}
\put(-7.2,3.5){\makebox(1,1)[t]{{\Large{(a)}}}}
\put(-7.2,2.0){\makebox(1,1)[t]{{\large{$K_{1}(x)$}}}}
\put(-3.5,-0.8){\makebox(1,1)[t]{{\large{$x$}}}}
\end{picture}

\vspace{-2.cm}

\begin{picture}(4,7)
\epsfysize=6cm
\epsfbox{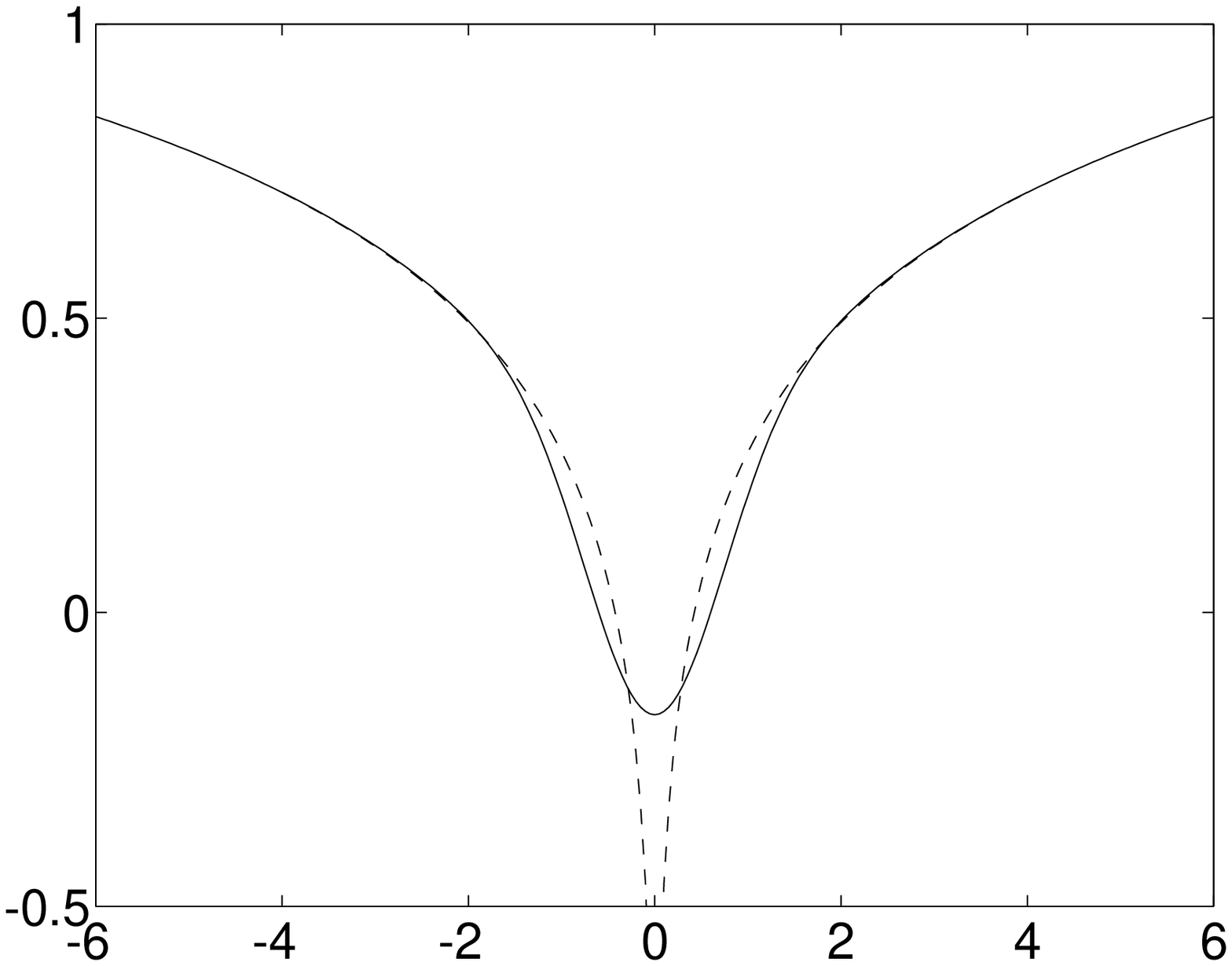}
\put(-7.2,3.5){\makebox(1,1)[t]{{\Large{(b)}}}}
\put(-7.2,2.0){\makebox(1,1)[t]{{\large{$K_{2}(x)$}}}}
\put(-3.5,-0.8){\makebox(1,1)[t]{{\large{$x$}}}}
\end{picture}

\vspace{-2.cm}

\begin{picture}(4,7)
\epsfysize=6cm
\epsfbox{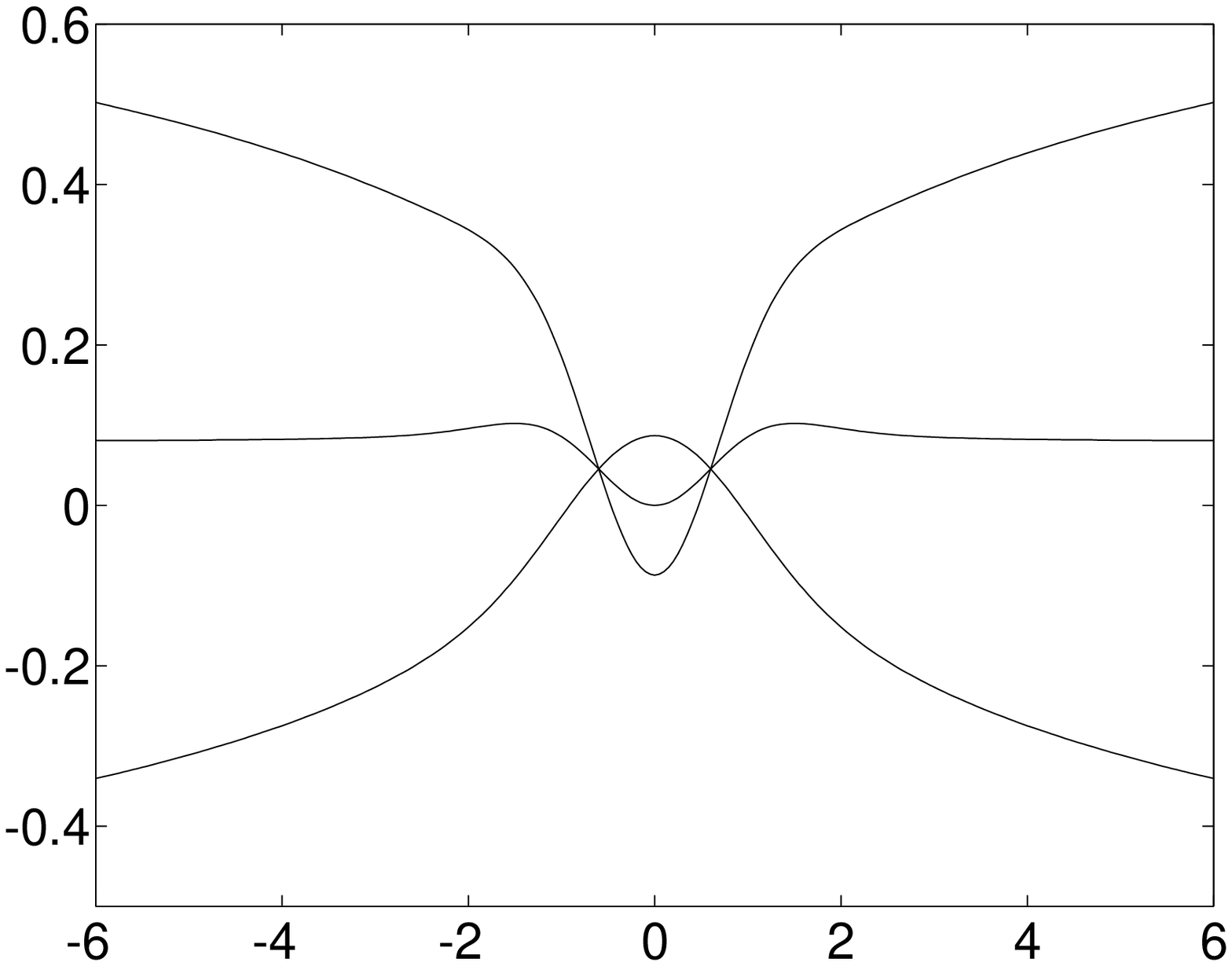}
\put(-7.2,3.5){\makebox(1,1)[t]{{\Large{(c)}}}}
\put(-7.2,2.0){\makebox(1,1)[t]{{\large{$K_{C^{2}}(x,\vartheta)$}}}}
\put(-3.5,-0.8){\makebox(1,1)[t]{{\large{$x$}}}}
\put(-2,3.4){\makebox(1,1)[t]{{\small{(1)}}}}
\put(-1.8,2){\makebox(1,1)[t]{{\small{(2)}}}}
\put(-2,0.8){\makebox(1,1)[t]{{\small{(3)}}}}
\end{picture}
\end{minipage}

\vspace{1.cm}

\caption{
The kernel for sampling the first (a) and second (b)
exponential-phase moment is shown (full line) and compared
with the classical result (dashed line). The kernel for
sampling $\langle \hat C^{2} \rangle$ (c) is shown for
$\vartheta$ $\!=$ $\!0$ (1), $\vartheta$ $\!=$ $\!\pi /4$ (2),
and $\vartheta$ $\!=$ $\!\pi /2$ (3).
\label{F1}
}
\end{figure}

The kernels $K_{1}(x)$ and $K_{2}(x)$ can be evaluated numerically
using standard routines. They are plotted in Fig.~\ref{F1}(a) and (b).
As expected, they rapidly approach the classical limits given in
Eqs.~(\ref{cl4}) and (\ref{cl5}) as $|x|$ increases.
Since they differ from the classical limits only in a small region
(of a few ``vacuum-fluctuation widths'') around zero, in practice
their evaluation needs applying Eqs.~(\ref{ker1}) and (\ref{ker2})
[or Eq.~(\ref{Ma5})] only for small values of $|x|$,
whereas for greater values the expressions given in Eqs.~(\ref{cl4})
and (\ref{cl5}) can be used. Note that for small values of $|x|$
power series expansion of the confluent hypergeometric function
in Eqs.~(\ref{ker1}) and (\ref{ker2}) can be used.

To verify URs connected with $\Delta \varphi$ [e.g., the
relations (\ref{u1}) and (\ref{02})], the first moment of $\hat E$
must be measured, which can be accomplished with the kernel
$K_{1}(x,\vartheta)$. With regard to URs of the type given in
Eqs.~(\ref{n3}) and (\ref{n3c}), one also needs the second moment of
$\hat E$ and the vacuum probability $\varrho_{00}$.
   From the above,
the kernels for sampling $\langle \hat C^{2} \rangle$ 
and $\langle \hat S^{2} \rangle$  read as
\begin{eqnarray}
\label{cos2}
K_{\pm}(x,\vartheta)\! = \!{\textstyle \frac{1}{4\pi}}
\big[1-\Phi (1,{\textstyle\frac{1}{2}},-x^{2})\big]
\!\pm\!{\textstyle \frac{1}{2}} \cos (2\vartheta) K_{2}(x)
\end{eqnarray}
[see Fig.~\ref{F1}(c)], where $K_{+}(x,\vartheta)$ $\!=$ 
$\!K_{C^{2}}(x,\vartheta)$ and $K_{-}(x,\vartheta)$ $\!=$ 
$\!K_{S^{2}}(x,\vartheta)$.

To conclude, we have presented a method for verification of number--phase 
URs. It is based on the possibility of direct sampling of exponential 
phase moments of the canonical phase of a single-mode quantum state, 
$\Psi_{k}$ $\!=$ $\!\langle\hat{E}^{k}\rangle$, from the data recorded 
in balanced homodyning. We have shown that the
corresponding kernels $K_{k}(x,\vartheta)$ are state independent and
well behaved. With increasing $|x|$ they rapidly approach the
classical limits. Since the method does not only apply to the
determination of low-order moments, it may also be used
for reconstructing the canonical phase distribution $p(\varphi)$
as a whole. However, a direct (state independent) sampling of $p(\varphi)$
according to Eq.~(\ref{n1}) seems to be impossible. If there would exist
a corresponding kernel, its Fourier components with respect to
$\vartheta$ would be equal to the kernels for determining $\Psi_{k}$.
Since for chosen $x$ the absolute values of these kernels increase
with $k$ [cf. Eqs.~(\ref{cl4}) and (\ref{cl5})], they cannot be treated
as Fourier coefficients of a well-behaved function of $\vartheta$.
Of course, this does not exclude an indirect reconstruction of
$p(\varphi)$. Measuring a limited number of $\Psi_{k}$, one can use,
e.g., the maximum entropy principle \cite{Jaynes} to obtain a
$p(\varphi)$ which best fits the measured values without
introducing any arbitrary bias. This also offers the possibility
of verification of URs that are not based on 
exponential phase moments \cite{Lindner,Birula}.

This work was supported by the Deutsche Forschungsgemeinschaft.


\end{document}